\documentclass{article}
\usepackage[T1]{fontenc}
\usepackage[utf8]{inputenc}
\usepackage{ismir} 
\usepackage{amsmath,cite,url}
\usepackage{graphicx}
\usepackage{color}
\usepackage{subcaption}
\usepackage{booktabs}
\usepackage[percent]{overpic}

\title{Emergent Tonal Structure in Learned Chord Embeddings and Its Relation to Tonal Tension}

\multauthor
  {Maral Ebrahimzadeh$^1$ \hspace{1cm} Gilberto Bernardes$^2$ \hspace{1cm} Sebastian Stober$^1$}
  {\\
  $^1$ Otto von Guericke University Magdeburg, Artificial Intelligence Lab, Magdeburg, Germany\\
  $^2$ INESC TEC, Faculty of Engineering, University of Porto, Portugal\\
  {\tt\small \{maral.ebrahimzadeh,stober\}@ovgu.de, gba@fe.up.pt}
  }

\def\authorname{M. Ebrahimzadeh, G. Bernardes, and S. Stober}

\begin{document}

\maketitle

\begin{abstract}
Several tonal pitch spaces and computational models have been proposed to analyze tonal structure in Western tonal music, many of them grounded in principles from music theory and used to support tonal analysis with important implications for tonal tension. In parallel, data-driven methods such as skip-gram have been used to learn chord embeddings from symbolic corpora, but their ability to recover tonal structure and its relation to tonal tension remains underexplored. In this work, we investigate how skip-gram chord embeddings reflect tonal structure and whether they provide a useful basis for analyzing structural aspects of tonal tension. Using chord sequences with and without transposition-based augmentation, we evaluate the learned spaces from geometric, functional, and tension-related perspectives. We show that augmented embeddings exhibit strong transposition equivariance, recover a clear circle-of-fifths structure, and support interpretable shifts between key-related regions of the learned space. We then derive embedding-based measures from chord-to-key distance and contextual chord-distance relations, and show that they capture meaningful aspects of tonal tension structure through correspondence with matched tonal measures and moderate alignment with human tension profiles. Across analyses, transposition-based augmentation generally improves the stability, tonal coherence, and interpretability of the learned space.
\end{abstract}

\section{Introduction}\label{sec:introduction}

The study of tonal structure has a long history in music theory, with roots in 18th-century harmonic theory~\cite{euler1739tentamen,rameau1971treatise}. Later, more explicit spatial and computational formulations emerged, including influential work by Longuet-Higgins~\cite{longuethiggins1962letters}. Tonal pitch spaces and related computational models have since been developed to characterize relationships among notes, chords, and keys within Western functional harmony~\cite{lerdahl1988tonal,chew2014mathematical,bernardes2016multilevel}. Such representations have been influential because they provide interpretable descriptions of tonal structure and support tasks including tonal similarity, key finding, and the analysis of tonal stability, expectation, and tension~\cite{chew2014mathematical,bernardes2016multilevel}. This spatial view is also central to computational accounts of tension, where perceived tension is often modeled in terms of distances and movements within a tonal space~\cite{navarrocaceres2020computational,herremans2019morpheus}.

At the same time, recent years have seen a growing interest in learning musical representations directly from data~\cite{choi2017tutorial,bretan2017learning}, including chord embeddings that capture harmonic context~\cite{chuan2020context,lazzari2022pitchclass2vec}. Skip-gram models are appealing because they learn from co-occurrence patterns alone~\cite{mikolov2013distributed}. Prior work has shown that skip-gram-based music embeddings can capture functional chord relationships, key relations, and an implicit circle-of-fifths structure in slice-based representations, suggesting that tonal structure can emerge from distributional learning even in relatively simple embedding spaces~\cite{chuan2020context}.

However, despite the importance of tonal spaces in music theory, cognition, and tonal tension research, the extent to which learned chord embedding spaces capture interpretable tonal structure remains underexplored. It is still unclear whether simple chord-level embeddings capture only local harmonic similarity or can also support richer tonal, transpositional, and tension-related structure.

We investigate this question using skip-gram chord embeddings trained on original and transposition-augmented chord sequences, and evaluate the resulting spaces through geometric, functional, and tension-related analyses. Code is publicly available.\footnote{\url{https://github.com/MaraalE/chord-tonality}}

\section{Related Work}

The representation of tonal structure in spatial form has a long tradition in music theory. Early geometric formulations by Longuet-Higgins~\cite{longuethiggins1962letters} laid the groundwork for later tonal space models, including Chew’s spiral array~\cite{chew2000towards,chew2014mathematical} and Lerdahl’s Tonal Pitch Space~\cite{lerdahl1988tonal, lerdahl2001tonal}, which provided interpretable geometries for relationships among pitches, chords, and keys. These models have also served as important foundations for computational tension modeling. Lerdahl later proposed a model of tonal tension grounded in Tonal Pitch Space~\cite{lerdahl2007modeling}, while later models such as MorpheuS~\cite{herremans2019morpheus} incorporated voice-leading information on top of spiral-array geometry. More recently, the Tonal Interval Space (TIS)~\cite{bernardes2016multilevel} provided another computational representation of tonal relations and was later used as the basis for a computational model of tonal tension~\cite{navarrocaceres2020computational}. Across these approaches, a common pattern is the use of hand-designed tonal geometries, often enriched with music-theoretic priors, as the basis for computational tension analysis.

In parallel, data-driven work has learned chord representations from symbolic corpora. Early work such as Chord2vec~\cite{madjiheurem2016chord2vec} applied skip-gram-style training to chord sequences, focusing primarily on training-objective comparisons rather than on detailed analysis of the learned geometry. Chuan et al.~\cite{chuan2020context} subsequently provided one of the first detailed analyses of distributional music embeddings, showing that embeddings of polyphonic slices capture functional chord and key relations and provide evidence of an implicit circle-of-fifths structure. Subsequent work extended learned chord and pitch-class representations to downstream tasks such as next-chord prediction and artist attribute prediction~\cite{lahnala2021chord} and symbolic structure segmentation~\cite{lazzari2022pitchclass2vec}, while more recent autoencoder and variational approaches~\cite{carvalho2023exploring,sadek2026circle} showed that circle-of-fifths geometry and key-related structure can also emerge in unsupervised latent spaces. Together, these studies suggest that tonal structure can emerge across different architectures and input granularities in learned representations~\cite{chuan2020context,lahnala2021chord,carvalho2023exploring,carvalho2024fourier,sadek2026circle}.

Our work is closest to Chuan et al.~\cite{chuan2020context}, but uses chord-token rather than polyphonic-slice embeddings, aligning the representation more directly with chord-based tonal analysis. More importantly, our goal is not only to show that tonal structure can emerge in a learned space, but to characterize how that structure behaves under transposition and how it supports downstream tonal analyses. In particular, we examine the effect of transposition-based augmentation, test transposition equivariance and stability explicitly, and evaluate whether the resulting geometry supports key-directed steering and interpretable tension-related probes.
This connects distributional chord embeddings with questions traditionally studied through hand-crafted tonal spaces and computational models of tension.

\section{Data and Training Setup}

\subsection{Datasets}

We use two annotated harmony datasets. Our primary dataset is the BPS-FH corpus~\cite{chen2018functional}, which provides expert Roman-numeral harmonic annotations for the first movements of Beethoven's 32 piano sonatas. We derive chord roots and qualities from these annotations using an adapted version of the released chord-symbol translation implementation~\cite{chen2019harmony}. As a secondary dataset, we use the Isophonics chord annotations~\cite{harte2005symbolic} for the Beatles catalogue, from which we use only the chord-label layer.

For each dataset, supported chord labels are reduced to a 24-token vocabulary of major and minor triads while preserving chord roots. Major-, minor-, and dominant-family labels, including sevenths and selected extensions, are mapped to their corresponding major or minor triads. Unsupported labels define sequence boundaries. This chord-label representation does not retain sevenths or inversions and does not encode voicing, limitations revisited in Section~\ref{sec:human_tension}. After event filtering and triad reduction, 9{,}542 and 13{,}470 major/minor chord events remain for BPS-FH and Isophonics, respectively. We then compress consecutive duplicates, remove sequences shorter than three tokens, and augment each corpus through all 12 pitch-class transpositions, balancing root distributions within each quality. Table~\ref{tab:corpus_stats} summarizes the resulting original corpora.

\begin{table}[!htbp]
\centering
\resizebox{\columnwidth}{!}{%
\begin{tabular}{|l|c|c|c|c|c|}
\hline
Dataset & Seqs. & Min & Max & Vocab & M/m cov. \\
\hline
BPS-FH     & 604 & 3 & 290 & 24 & 89.1\% \\
\hline
Isophonics & 515 & 3 & 128 & 24 & 94.9\% \\
\hline
\end{tabular}%
}
\caption{Summary statistics of the original skip-gram corpora, including the number of sequences, sequence-length range, vocabulary size, and the proportion of valid chord events retained after triad reduction and filtering.}
\label{tab:corpus_stats}
\end{table}

\subsection{Embedding Model and Training Setup}

We learn chord embeddings with a skip-gram model implemented in \texttt{gensim}~\cite{rehurek_lrec}, using it as a minimal test of whether tonal and tension-related structure can emerge from chord co-occurrence statistics alone. A simple model provides a cleaner test of this hypothesis than a more complex architecture. Each chord is represented as a discrete \texttt{root\_quality} token, such as \texttt{C\_M} or \texttt{A\_m}, and the model predicts surrounding context tokens from a center token. All models use a fixed context window of 5 to capture short-range harmonic context, negative sampling with 5 negatives, \texttt{min\_count}=1, no subsampling, and 64-dimensional embeddings, at the lower end of the range used in prior chord-embedding work~\cite{lahnala2021chord,chuan2020context}. Prior work found embedding dimension to have only a minor effect~\cite{lahnala2021chord}. Training is single-threaded and uses fixed random seeds.

We compare two training settings. In the \textsc{Augmented} setting, embeddings are learned from the 12-fold transposed corpus for 50 epochs. In the \textsc{Original} setting, they are learned from the non-augmented corpus for 600 epochs, approximately matching total training-pair exposure despite the smaller corpus. For each setting, we train two random initializations and, unless otherwise noted, report metrics averaged over the two seeds.

\section{Tonal Structure in Learned Chord Embeddings}
\subsection{Circle-of-Fifths Structure}

\begin{figure*}[t]
    \centering

    \begin{subfigure}{0.48\textwidth}
        \centering
        \includegraphics[width=\linewidth]{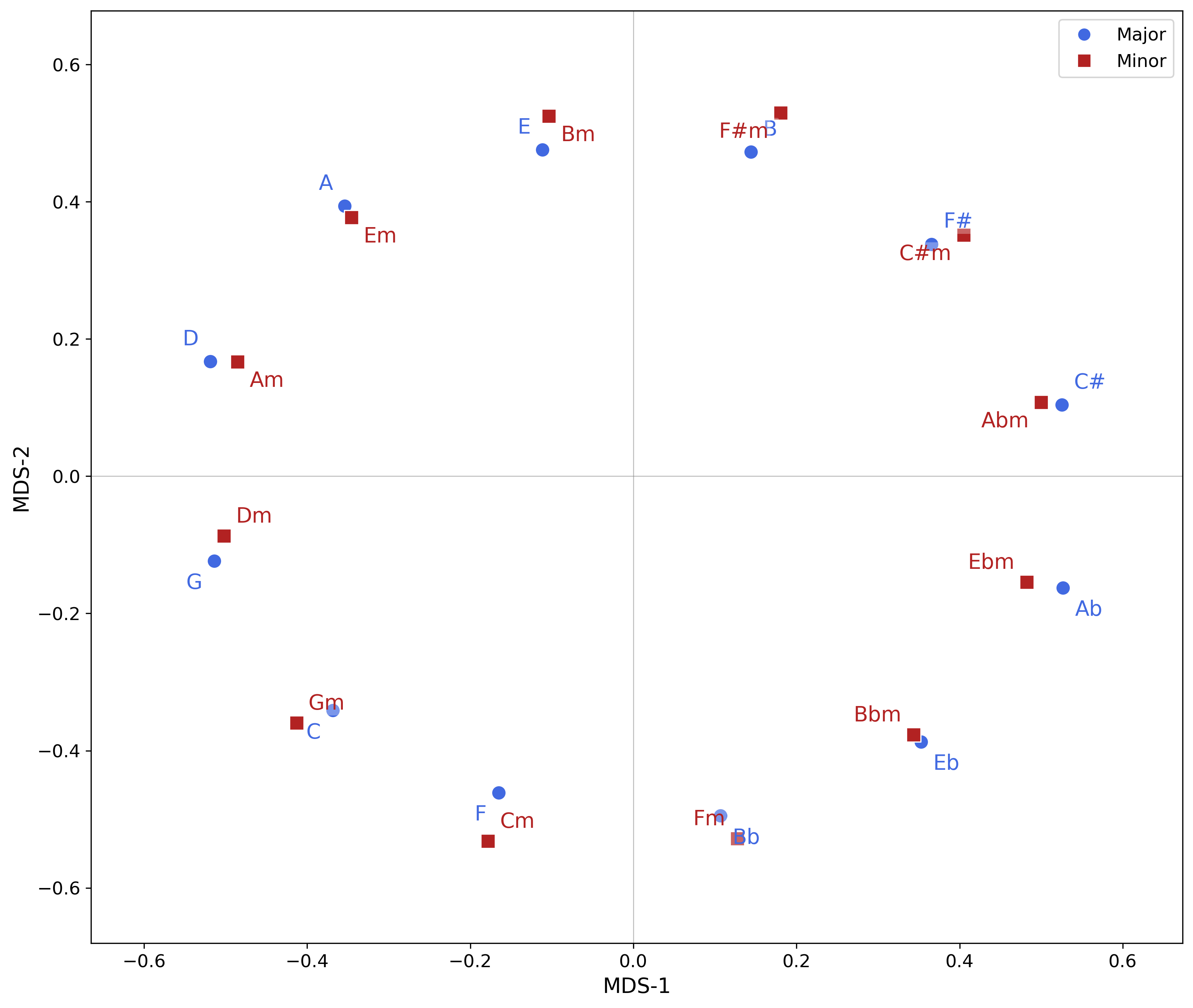}
        \caption{Augmented}
        \label{fig:cof_augmented}
    \end{subfigure}
    \hfill
    \begin{subfigure}{0.48\textwidth}
        \centering
        \includegraphics[width=\linewidth]{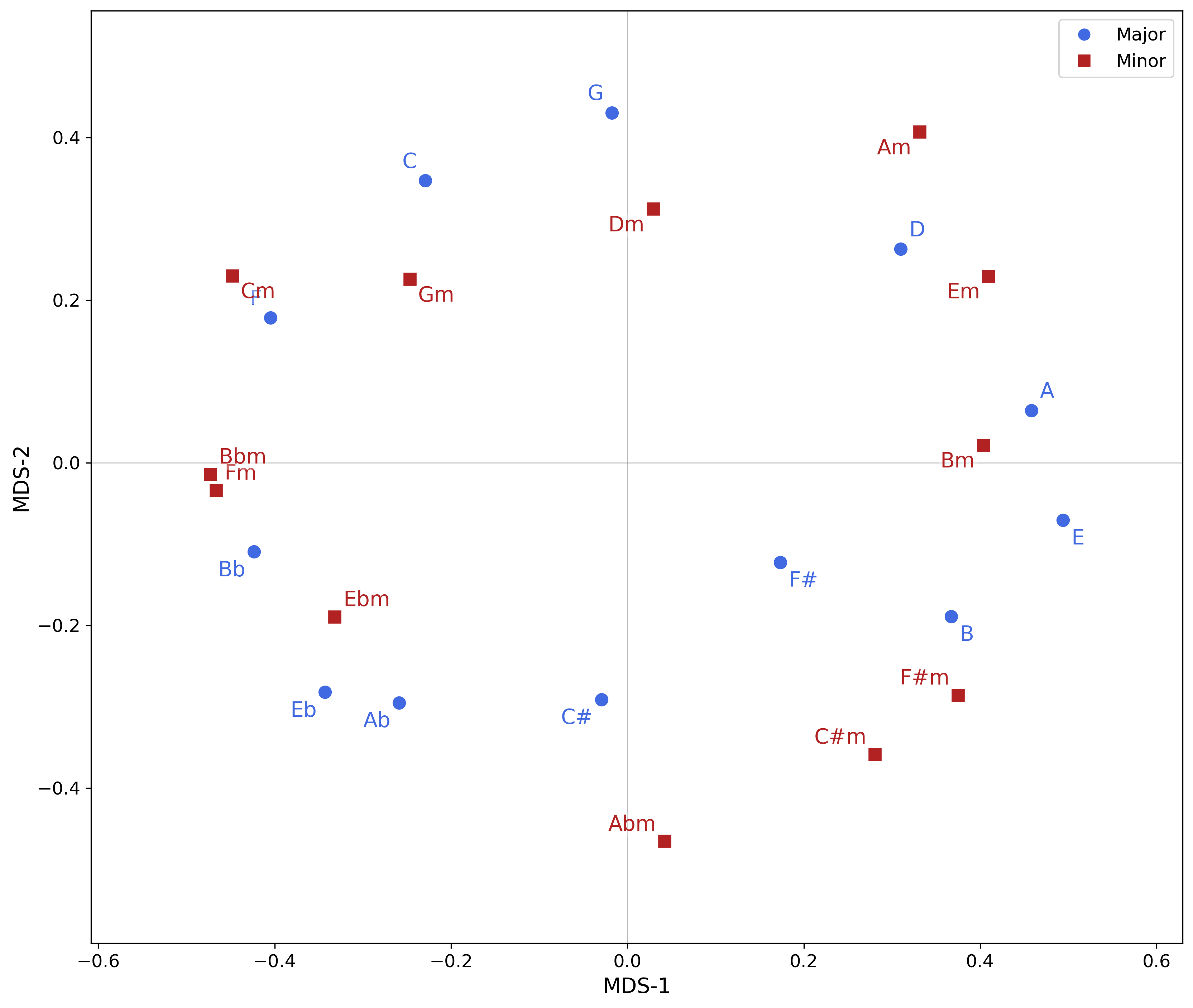}
        \caption{Non-augmented}
        \label{fig:cof_non_augmented}
    \end{subfigure}

    \caption{Two-dimensional MDS projections of the learned 24-chord spaces for representative augmented and non-augmented models.}
    \label{fig:cof_geometry}
\end{figure*}

The circle of fifths is a canonical model of tonal proximity in Western harmony. We test whether this organization emerges from chord co-occurrence statistics in the learned embedding spaces.
Figure~\ref{fig:cof_geometry} shows two-dimensional projections of the learned 24-chord spaces for representative \textsc{Augmented} (AUG) and \textsc{Original} (ORIG) models, obtained with classical MDS, which seeks to preserve global pairwise distances and is well suited to inspecting cyclic structure in a low-dimensional point set. In the augmented model, the learned geometry more clearly reflects circle-of-fifths structure, with major and minor regions occupying more systematically aligned positions. The non-augmented model still exhibits tonal organization, but the arrangement is less regular, particularly in the minor space.

To quantify this structure, we use three complementary measures. Cyclic ordering error evaluates how closely the angular ordering of chords in the MDS projection matches the canonical circle-of-fifths order up to rotation and reversal. Template correlation ($\rho$) measures the Spearman correlation between learned pairwise distances and an idealized circle-of-fifths distance template. Finally, Fifth R@2 measures how often the two fifth-related neighbors of a chord are recovered among its two nearest neighbors within the same chord quality.

Table~\ref{tab:cof_metrics} shows that the augmented models generally exhibit stronger circle-of-fifths structure than the non-augmented comparison models on both datasets. On BPS-FH, they achieve perfect cyclic ordering for both chord qualities, and Isophonics shows the same overall pattern, particularly in the minor-quality space.

\begin{table}[!htbp]
\centering
\resizebox{\columnwidth}{!}{%
\begin{tabular}{|l|l|c|c|c|}
\hline
Dataset & Model & Cyclic error $\downarrow$ & Template $\rho \uparrow$ & Fifth R@2 $\uparrow$ \\
\hline
BPS-FH     & AUG  & 0.000 / 0.000 & 0.868 / 0.958 & 1.000 / 1.000 \\
\hline
BPS-FH     & ORIG & 0.007 / 0.007 & 0.736 / 0.793 & 0.979 / 0.771 \\
\hline
Isophonics & AUG  & 0.000 / 0.000 & 0.914 / 0.780 & 1.000 / 0.958 \\
\hline
Isophonics & ORIG & 0.000 / 0.111 & 0.769 / 0.594 & 0.729 / 0.750 \\
\hline
\end{tabular}%
}
\caption{Circle-of-fifths structure (major / minor). Lower cyclic error and higher template $\rho$ and Fifth R@2 indicate stronger structure.}
\label{tab:cof_metrics}
\end{table}

\subsection{Transposition Equivariance}

A central motivation for transposition-based augmentation is that harmonically similar patterns should induce comparable structure across keys. We test whether transposing the chord vocabulary by $k$ semitones approximately preserves embedding structure using two measures: Procrustes alignment error between the original embedding configuration and its transposed counterpart, and neighbor overlap measured by Jaccard similarity between transposed top-8 nearest-neighbor sets. Lower Procrustes error indicates stronger global equivariance, while higher Jaccard indicates stronger local consistency.

Table~\ref{tab:equivariance_metrics} summarizes the mean scores over non-identity transpositions for both datasets. Across BPS-FH and Isophonics, the augmented models achieve lower Procrustes error and higher neighbor overlap than the non-augmented models, indicating stronger global alignment and local consistency under transposition. Because every sequence is transposed through all 12 pitch-class shifts, the augmented training statistics are transposition-symmetric by construction, so this consistency is partly built into the data. More informatively, the ORIG models retain partial local consistency despite weak global Procrustes alignment, indicating that some transposition structure emerges from co-occurrence alone, while augmentation makes it more precise and consistent.

\begin{table}[!htbp]
\centering
\resizebox{\columnwidth}{!}{%
\begin{tabular}{|l|l|c|c|}
\hline
Dataset & Model & Procrustes $\downarrow$ & Jaccard $\uparrow$ \\
\hline
BPS-FH     & AUG  & 0.074 & 0.875 \\
\hline
BPS-FH     & ORIG & 0.464 & 0.609 \\
\hline
Isophonics & AUG  & 0.180 & 0.792 \\
\hline
Isophonics & ORIG & 0.622 & 0.480 \\
\hline
\end{tabular}%
}
\caption{Mean transposition-equivariance scores over the 11 non-identity
pitch-class shifts.}
\label{tab:equivariance_metrics}
\end{table}

\subsection{Emergent Key Structure}

Beyond global geometry, we ask whether the learned space organizes chords around key-centered tonal regions. In tonal pitch spaces, chords belonging to the same diatonic collection are expected to occupy a relatively coherent local region associated with the corresponding key. To test whether such structure emerges in the learned embeddings, we represent each key by the mean vector of a key-specific chord set, which we refer to as a key centroid for convenience.
For example, the centroid for C major is computed from the embeddings of its diatonic triads \texttt{C\_M}, \texttt{D\_m}, \texttt{E\_m}, \texttt{F\_M}, \texttt{G\_M}, and \texttt{A\_m}.
For minor keys, we use a functional chord set in which the dominant is major. For example, the set for A minor is \texttt{A\_m}, \texttt{C\_M}, \texttt{D\_m}, \texttt{E\_M}, \texttt{F\_M}, and \texttt{G\_M}.
We then compare the distances of in-key and out-of-key chords to these key centroids. To avoid circularity, in-key distances are computed in a leave-one-out
manner, so that each chord is evaluated against a centroid formed from
the remaining chords in that key-specific set.

Table~\ref{tab:key_structure_metrics} summarizes two complementary measures of this key-centered structure: the leave-one-out gap between out-of-key and in-key distances, and the corresponding AUC measure $P(d_{\text{in}} < d_{\text{out}})$. On both BPS-FH and Isophonics, the augmented models show stronger tonal separation than the non-augmented models in major and minor keys, with especially clear gains in the minor condition.

\begin{table}[!htbp]
\centering
\resizebox{\columnwidth}{!}{%
\begin{tabular}{|l|l|c|c|}
\hline
Dataset & Model & LOO gap $\uparrow$ & AUC $\uparrow$ \\
\hline
BPS-FH     & AUG  & 0.363 / 0.313 & 0.931 / 0.891 \\
\hline
BPS-FH     & ORIG & 0.281 / 0.239 & 0.891 / 0.849 \\
\hline
Isophonics & AUG  & 0.369 / 0.268 & 0.946 / 0.857 \\
\hline
Isophonics & ORIG & 0.269 / 0.197 & 0.922 / 0.825 \\
\hline
\end{tabular}%
}
\caption{Summary of emergent key structure. Values are reported as major / minor. The leave-one-out (LOO) gap is defined as $\mathrm{mean}(d_{\text{out}}) - \mathrm{mean}(d_{\text{in}})$, so higher values indicate stronger separation between out-of-key and in-key chords. Higher AUC indicates better discrimination of in-key versus out-of-key distances.}
\label{tab:key_structure_metrics}
\end{table}

Overall, the augmented embeddings recover not only circular tonal proximity, but also stronger and more consistent key-centered structure across datasets. BPS-FH controls support this finding. Random and label-permuted embeddings fail to recover key structure, with key-discrimination AUC near or below chance ($\approx 0.5$ versus $0.89$--$0.93$ for AUG). Global shuffling largely collapses tonal separation and equivariance (AUC $0.46/0.58$, Procrustes error $0.84$ versus $0.07$), whereas within-sequence shuffling preserves circle-of-fifths ordering and most key separation (AUC $\approx 0.92$) but increases Procrustes error ($0.17$ versus $0.07$). This suggests that broad tonal structure depends more on sequence-level co-occurrence than exact chord order.

\subsection{Tonal Region Steering}
\label{sec:steering}

To test whether differences between key centroids define interpretable directions in the learned space, we construct steering vectors from a source-key centroid to a target-key centroid and apply them to source-only chords, i.e., chords that belong to the source-key set but not to the target-key set. We evaluate steering using three measures: continuous preference change $\Delta$, defined as the post-steering minus pre-steering change in $d(x,\text{tgt}) - d(x,\text{src})$ (cosine distance $d$), target-hit rate, and source-escape rate. More negative $\Delta$ indicates a stronger shift toward the target region. Target-hit rate measures how often the nearest chord after steering belongs to the target-key set, while source-escape rate measures how often the steered chord leaves the source-key set.
Table~\ref{tab:steering_metrics} reports results at steering strength $\alpha=2.0$. On both datasets, the augmented models produce larger shifts toward the target region and higher target-hit and source-escape rates than the non-augmented models, indicating a more coherent key-related geometry.

\begin{table}[!htbp]
\centering
\resizebox{\columnwidth}{!}{%
\begin{tabular}{|l|l|c|c|c|}
\hline
Dataset & Model & Cont.\ $\Delta \downarrow$ & Target hit $\uparrow$ & Source escape $\uparrow$ \\
\hline
BPS-FH & AUG  & -0.921 & 0.630 & 0.685 \\
\hline
BPS-FH & ORIG & -0.724 & 0.377 & 0.410 \\
\hline
Isophonics & AUG  & -0.840 & 0.639 & 0.667 \\
\hline
Isophonics & ORIG & -0.678 & 0.295 & 0.305 \\
\hline
\end{tabular}%
}
\caption{Tonal region steering results on source-only chords at $\alpha=2.0$.}
\label{tab:steering_metrics}
\end{table}

\section{Relation to Tonal Tension}

Tonal-tension models typically define tension in terms of harmonic displacement in tonal space, combining local chord-to-chord relations and distance from a prevailing key center. We ask whether analogous quantities can be defined in the learned embedding space, examining transposition stability (Section~\ref{sec:tension_stability}), correspondence with matched Tonal Interval Vector (TIV) quantities (Section~\ref{sec:tiv_relation}), and relation to human tension ratings (Section~\ref{sec:human_tension}).

\subsection{Tension Measures and Transposition Stability}
\label{sec:tension_stability}

We begin by defining sequence-level tension measures directly from distances to tonal context in the embedding space. Our main measure, given in Eq.~\ref{eq:tkey}, computes at each chord position $t$ the expected distance between the current chord embedding and the centroids of candidate keys, weighted by a prefix-based soft distribution over keys. Intuitively, this quantity is low when the current chord is well aligned with the inferred tonal context and high when it is distant from the likely key structure.
Formally, let $c_t$ denote the embedding of the chord at position $t$, let $\mu_K$ denote the centroid of the triads associated with key $K$, and let $\mu^{\mathrm{LOO}}_K(t)$ denote the corresponding centroid after excluding the chord at position $t$ when it belongs to that key's triadic collection. We estimate the soft key distribution as $p_t(K)=\operatorname{softmax}_K\!\left( \cos(\bar{c}_{1:\min(t,8)},\mu_K)/0.25\right)$, where $\bar{c}_{1:\min(t,8)}$ is the mean embedding of the chord prefix, capped at the first eight chords. For this analysis, minor-key centroids use the natural-minor triadic collection. Using cosine distance $d(\cdot,\cdot)$, we define
\begin{equation}
T_{\mathrm{key}}(t)=\sum_K p_t(K)\, d\!\left(c_t,\mu^{\mathrm{LOO}}_K(t)\right).
\label{eq:tkey}
\end{equation}

To capture short-range harmonic change, we also define a contextual chord-distance term,
\begin{equation}
T_{\mathrm{ctx}}(t)=d\!\left(c_t,\bar{c}_{t-W:t-1}\right),
\label{eq:tctx}
\end{equation}
where $\bar{c}_{t-W:t-1}$ is the centroid of the previous $W=3$ chord embeddings. Finally, we report a simple combined variant that augments the key-relative term with local contextual distance:
\begin{equation}
T_{\mathrm{comb}}(t)=T_{\mathrm{key}}(t)+\beta\,T_{\mathrm{ctx}}(t),
\qquad \beta = 0.5.
\label{eq:tcomb}
\end{equation}

A basic requirement for a musically meaningful tonal representation is that its induced tension profile should remain stable under transposition. We therefore evaluate each measure by computing its tension curve for a chord sequence, transposing the sequence by $k \in \{0,\dots,11\}$ semitones, recomputing the curve, and comparing the original and transposed curves using relative $\ell_2$ error and correlation. All models are evaluated on the same model-independent subset of sequences containing at least six chords. Table~\ref{tab:tension_stability_main} summarizes the results for the non-identity transpositions $k=1,\dots,11$.

\begin{table}[!htbp]
\centering
\setlength{\tabcolsep}{3.5pt}
\small
\resizebox{\columnwidth}{!}{%
\begin{tabular}{|l|l|c|c|c|c|c|c|}
\hline
Dataset & Model
& \multicolumn{2}{c|}{$T_{\mathrm{key}}$}
& \multicolumn{2}{c|}{$T_{\mathrm{ctx}}$}
& \multicolumn{2}{c|}{$T_{\mathrm{comb}}$ ($\beta=0.5$)} \\
\hline
&
& relL2 $\downarrow$ & corr $\uparrow$
& relL2 $\downarrow$ & corr $\uparrow$
& relL2 $\downarrow$ & corr $\uparrow$ \\
\hline
BPS-FH     & AUG  & 0.063 & 0.883 & 0.074 & 0.995 & 0.065 & 0.976 \\
\hline
BPS-FH     & ORIG & 0.173 & 0.321 & 0.246 & 0.925 & 0.180 & 0.770 \\
\hline
Isophonics & AUG  & 0.144 & 0.804 & 0.200 & 0.968 & 0.155 & 0.914 \\
\hline
Isophonics & ORIG & 0.246 & 0.305 & 0.391 & 0.861 & 0.267 & 0.696 \\
\hline
\end{tabular}%
}
\caption{Transposition stability averaged over sequences and non-identity transpositions ($k=1,\dots,11$). Lower relative $\ell_2$ is better, higher correlation is better.}
\label{tab:tension_stability_main}
\end{table}

The strongest contrast appears for $T_{\mathrm{key}}$. On both datasets, the augmented models are substantially more stable than the non-augmented models, showing that augmentation better preserves key-relative tension under pitch-class shifts.
The contextual chord-distance measure $T_{\mathrm{ctx}}$ retains much higher curve-shape correlation under transposition, especially on BPS-FH, whereas the combined measure $T_{\mathrm{comb}}$ lies between the two extremes. Thus, augmentation primarily strengthens global key-relative organization, whereas short-range contextual structure remains relatively robust without it.

\subsection{Relation to TIV-Based Tonal Tension Components}
\label{sec:tiv_relation}

We next compare the local quantities underlying the embedding-based measures with matched quantities derived from TIVs~\cite{bernardes2016multilevel}, a 12-dimensional Fourier-based representation of pitch-class collections widely used for modeling tonal relations. The contextual chord-distance term is intended to reflect short-range harmonic displacement between successive or recent chords, which in music-theoretic terms is closely related to common-tone relations. We therefore compare it with a matched TIV-based contextual phase-distance quantity. Likewise, we compare the embedding-based key-relative chord-distance term with a TIV-based phase-distance quantity relative to the inferred key representation. For the key-relative comparison, both representations use the same hard key estimate $\hat{K}_t=\arg\max_K p_t(K)$. Rather than correlating full tension curves directly, we compare these matched local components on the same chord sequences. Table~\ref{tab:tiv_matched} summarizes the resulting correlations on both BPS-FH and Isophonics.

\begin{table}[!htbp]
\begin{center}
\resizebox{\columnwidth}{!}{%
\begin{tabular}{|l|l|c|c|c|c|}
\hline
& & \multicolumn{2}{c|}{Contextual chord distance} & \multicolumn{2}{c|}{Key-relative chord distance} \\
\cline{3-6}
Dataset & Model & Pearson $\uparrow$ & Spearman $\uparrow$ & Pearson $\uparrow$ & Spearman $\uparrow$ \\
\hline
BPS-FH & AUG  & 0.733 & 0.701 & 0.624 & 0.698 \\
\hline
BPS-FH & ORIG & 0.707 & 0.687 & 0.513 & 0.511 \\
\hline
Isophonics & AUG  & 0.637 & 0.628 & 0.526 & 0.592 \\
\hline
Isophonics & ORIG & 0.617 & 0.599 & 0.422 & 0.440 \\
\hline
\end{tabular}
}
\end{center}
\caption{Length-weighted means of per-sequence correlations between embedding and matched TIV-based quantities. Higher values indicate stronger correspondence.}
\label{tab:tiv_matched}
\end{table}

On BPS-FH, the contextual distance measure shows clear positive correspondence with its matched TIV-based quantity in both model families, with Pearson correlations around 0.71--0.73 and Spearman correlations around 0.69--0.70. This indicates that both augmented and non-augmented embeddings capture local harmonic relations in a way that is broadly consistent with the matched TIV-based proxy. The clearest difference appears in the key-relative comparison, where the augmented BPS-FH models show substantially stronger correspondence than the non-augmented models in both Pearson and Spearman correlation. The same qualitative pattern is also observed on Isophonics: while the contextual term remains relatively similar across model families, the augmented models again show a clearer advantage on the key-relative term.

Overall, the embeddings recover harmonic organization consistent with matched TIV quantities, and the larger key-relative gains suggest that augmentation primarily strengthens the tonal organization of the learned space.

\subsection{Relation to Human Tension Ratings}
\label{sec:human_tension}

Finally, we test whether a lightweight dynamic measure built from the learned embedding space relates to human tension ratings on BPS-FH, using the same 12 short chord progressions and their mean human tension ratings from Navarro-Cáceres et al.~\cite{navarrocaceres2020computational}. Our goal is not to outperform earlier computational tension models, but to assess how much tension-related structure can be recovered from the learned embedding space.

\begin{figure*}[!htbp]
    \centering
    \includegraphics[width=0.96\textwidth]{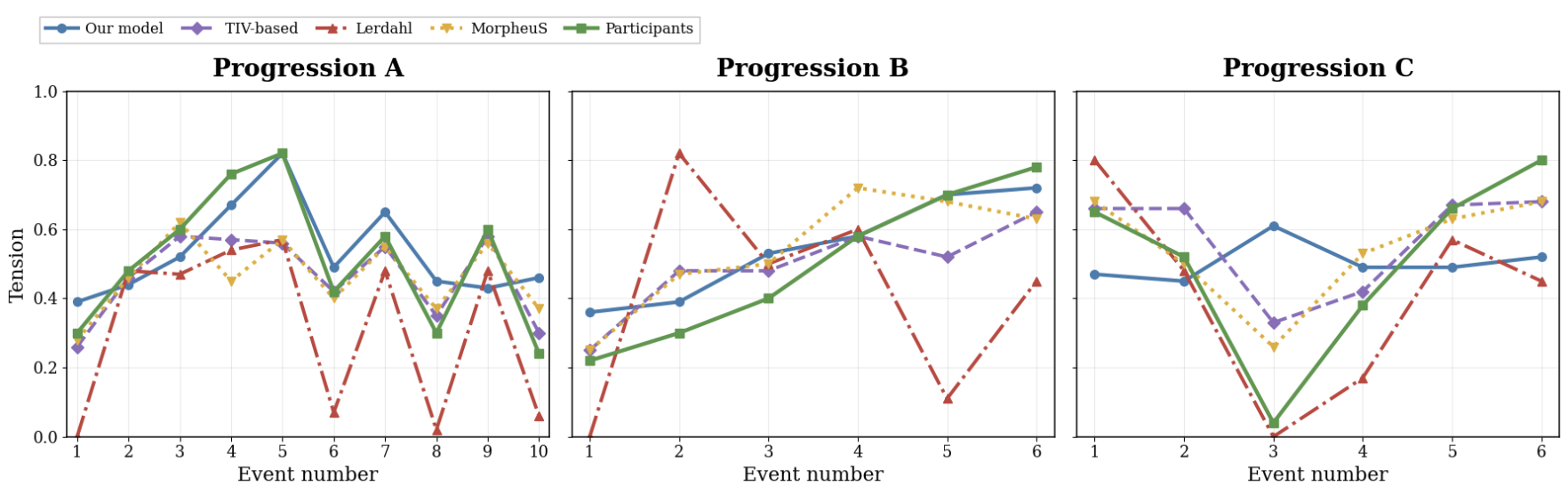}
    \caption{Predicted and human-rated tension curves for three representative progressions, comparing the embedding-based measure with TIV, Lerdahl’s model, MorpheuS, and mean participant ratings.}
    \label{fig:human_tension_examples}
\end{figure*}

For this analysis, we use a single representative augmented model (seed~1). To construct this measure, we combine two components derived from the embedding: a contextual chord-distance term (Eq.~\ref{eq:tctx}) and a key-relative chord-distance term (Eq.~\ref{eq:tkey}). In addition, we include two hand-designed terms: a repetition term and a simple local resolution term defined from decreases in key-relative distance and repetition. These quantities are z-normalized, combined linearly, smoothed with a first-order memory process, and mapped to the unit interval, yielding a lightweight dynamic measure that can be compared directly to the human ratings.

We estimate the combination weights by fitting the dynamic measure to the mean human ratings across all 12 progressions. In this all-12 fit, the largest weight is assigned to the key-relative term ($w_{\mathrm{key}}=0.491$), followed by the resolution term ($w_{\mathrm{res}}=0.343$), with smaller contributions from repetition ($w_{\mathrm{rep}}=0.108$) and contextual chord distance ($w_{\mathrm{ctx}}=0.057$), suggesting that the measure is driven primarily by key-related harmonic fit and local release.
On the full set of 12 progressions, the measure achieves moderate correspondence with the mean human ratings ($r=0.473$, $\rho=0.531$, RMSE$=0.189$). Under leave-one-progression-out evaluation, the model and normalization statistics are fit on 11 progressions and applied to the held-out one. Performance decreases only moderately ($r=0.406$, $\rho=0.456$, RMSE$=0.199$), suggesting that the relationship is not purely idiosyncratic to the fitted set.
Per-progression analysis shows that the proposed measure reaches strong correlation ($r \ge 0.6$) on 7 of 12 progressions in both the all-12 fit and LOOCV. The lower overall correlation is driven mainly by a smaller subset of cases, such as Progression~C, where tension depends on factors not encoded in chord labels.

\begin{table}[!htbp]
\centering
\begin{tabular}{|l|c|c|c|}
\hline
Setting & Pearson $r$ & Spearman $\rho$ & RMSE \\
\hline
All-12 fit & 0.473 & 0.531 & 0.189 \\
\hline
LOOCV & 0.406 & 0.456 & 0.199 \\
\hline
\end{tabular}
\caption{Global performance of the embedding-based dynamic measure against mean human tension ratings on the 12 progressions.}
\label{tab:human_tension_global}
\end{table}

\begin{figure}[t]
    \centering

    \begin{overpic}[width=\columnwidth]{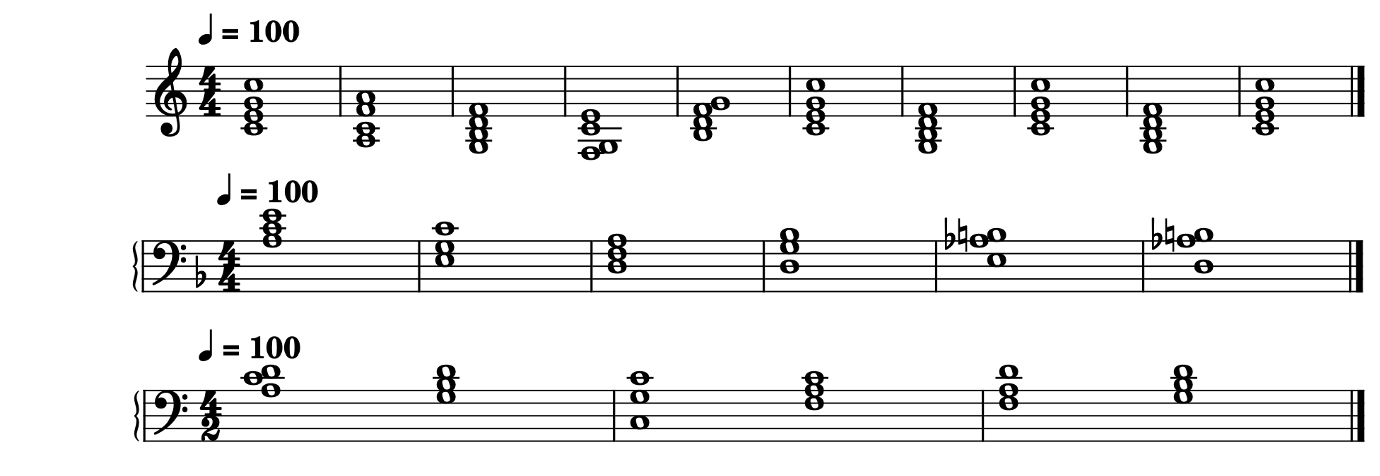}
        \put(1,27){\large\textbf{(a)}}
        \put(1,14){\large\textbf{(b)}}
        \put(1,3){\large\textbf{(c)}}
    \end{overpic}

    \caption{Score representations of the three representative examples discussed in Figure~\ref{fig:human_tension_examples}: (a) Progression~A, (b) Progression~B, and (c) Progression~C.}
    \label{fig:progression_scores}
\end{figure}

Figure~\ref{fig:human_tension_examples} shows three representative examples. Progression~A is a consensus case in which all models broadly recover the human rise-and-fall profile, suggesting that its tension curve is driven by salient harmonic transitions accessible to multiple tonal representations. Progression~B is especially encouraging for the proposed approach, as the embedding-based measure closely follows the participant ratings. By contrast, Progression~C highlights a limitation of the method: the model fails to reproduce a pronounced change in the human ratings, likely because such judgments depend on musical factors not encoded in chord labels alone, including inversion.

Overall, the measure captures a meaningful subset of tonal tension while clarifying the limits of chord-label embeddings. Because the input omits inversion, realized bass motion, voicing, rhythmic salience, and non-chord tones, we interpret this experiment as diagnostic rather than a full predictor of human tension ratings.

\section{Conclusion and Future Work}

We showed that simple skip-gram chord embeddings can serve as interpretable tonal spaces, recovering circle-of-fifths organization, transposition equivariance, key-centered separation, tonal-region steering, and tension-related structure. These properties were generally stronger with transposition augmentation, highlighting its central role in producing clearer and more stable harmonic organization.

Embedding distances correspond to matched TIV quantities, and a lightweight dynamic measure shows moderate agreement with human ratings. Similar trends across BPS-FH and Isophonics suggest that the findings are not confined to one repertoire, although differences in magnitude may reflect corpus and annotation differences rather than genre alone. We therefore view the tension analysis as diagnostic rather than a complete perceptual model. Future work should evaluate larger heterogeneous corpora such as ChoCo~\cite{deberardinis2023choco}, expand the chord vocabulary beyond triads, and incorporate contextualized or note-level information.

\section{AI Usage Statement}
An AI-based language assistant was used to support text revision and wording refinement during manuscript preparation.

\bibliography{ISMIR}
\end{document}